\documentclass[conference]{IEEEtran}
\usepackage{array, calc, color, multicol}
\usepackage{algorithm}
\usepackage[noend]{algorithmic} %, algorithm2e}
\usepackage{graphics, epstopdf, graphicx, epsfig, psfrag, epsf, subfigure}
\usepackage{amssymb, amsfonts, amsmath, times}
\usepackage{relsize}
\usepackage{multicol,balance, verbatim}
\usepackage{textcomp, mathrsfs, stmaryrd, setspace}
\usepackage{amssymb}
\newcommand{\ie}{{\em i.e., }}
\newcommand{\Ie}{{\em I.e., }}
\newcommand{\eg}{{\em e.g., }}
\newcommand{\Eg}{{\em E.g., }}

\newtheorem{example}{Example}

\DeclareGraphicsExtensions{.eps, .pdf, .png, .jpg}   % optional

\newcommand{\Sset}{\mathcal{S}}

\newcommand{\Nset}{\mathcal{N}}
\newcommand{\Mset}{\mathcal{M}}
\newcommand{\Hset}{\mathcal{H}}

\newcommand{\Lset}{\mathcal{L}}

\newcommand{\Wset}{\mathcal{W}}

\begin{document}
%\title{Prioritization in IDNC for Completion Time Reduction or Distortion Minimization}
%\title{On the Performance Bounds of Hybrid-interface Cooperative Network}
\title{Content-Aware Network Coding over Device-to-Device Networks}

%On the Performance of Network Coding for Cooperative Mobile Devices with Multiple Interfaces
%On the Minimum Number of Transmissions in Single-Hop Wireless Coding Networks
%On coding for cooperative data exchange

%\author{Yasaman Keshtkarjahromi, Hulya Seferoglu, Rashid Ansari\\
%{\small ECE Department, University of Illinois at Chicago}\\
%{ \small \tt ykesht2@uic.edu, hulya@uic.edu, ransari@uic.edu}\\
%\and
%Ashfaq Khokhar\\
%{\small ECE Department, Illinois Institute of Technology}\\
%{ \small \tt ashfaq@iit.edu }\\
%}

\author{Yasaman Keshtkarjahromi, Hulya Seferoglu, Rashid Ansari\\
{\small University of Illinois at Chicago}\\
{ \small \tt ykesht2@uic.edu, hulya@uic.edu, ransari@uic.edu}\\
\and
Ashfaq Khokhar\\
{\small Illinois Institute of Technology}\\
{ \small \tt ashfaq@iit.edu }\\
}

\maketitle

{$\hphantom{a}$}\vspace{-30pt}{}

\allowdisplaybreaks

\begin{abstract}
Consider a scenario of broadcasting a common content to a group of cooperating mobile devices that are within proximity of each other.
Devices in this group may receive partial content from the source due to packet losses over wireless broadcast links. We further consider that packet losses are different for different devices.
The remaining missing content at each device can then be recovered, thanks to cooperation among the devices by exploiting device-to-device (D2D) connections. In this context, the minimum amount of time that can guarantee a complete acquisition of the common content at every device is referred to as the ``completion time''. It has been shown that instantly decodable network coding (IDNC) reduces the completion time as compared to no network coding in this scenario.
Yet, for applications such as video streaming, not all packets have the same importance and not all devices are interested in the same quality of content. This problem is even more interesting when additional, but realistic constraints, such as strict deadline, bandwidth, or limited energy are added in the problem formulation. We assert that direct application of IDNC in such a scenario yields poor performance in terms of content quality and completion time. In this paper, we propose a novel Content and Loss-Aware IDNC scheme that improves content quality and network coding opportunities jointly by taking into account importance of each packet towards the desired quality of service (QoS) as well as the channel losses over D2D links.
Our proposed Content and Loss-Aware IDNC (i) maximizes the quality under the completion time constraint, and (ii) minimizes the completion time under the quality constraint. We demonstrate the benefits of Content and Loss-Aware IDNC through simulations.
\end{abstract}


\maketitle

%\IEEEdisplaynontitleabstractindextext

%\IEEEpeerreviewmaketitle

\section{Introduction}\label{sec:introduction}
\IEEEPARstart{T}{he} widely-used and popular applications in today's mobile devices come with increasing demand for high quality content, bandwidth, and energy \cite{cisco_rep}, \cite{ericsson_rep}. Cooperation among mobile devices, facilitated by improved computational, storage, and connectivity capabilities of these devices, is a promising approach to meet these demands.

In this paper, we consider an increasingly popular application of broadcasting a common content (\eg video), to a group of cooperating mobile devices within proximity and transmission range of each other. \Eg a group of friends may be interested in watching the same video on YouTube, or a number of students may participate in an online education class. In such a scenario, the content server may just broadcast the video via cellular links. However, mobile devices may receive only a partial content due to packet losses over wireless broadcast links. The remaining missing content can then be recovered thanks to cooperation among the devices via device-to-device (D2D) connections  such as WiFi-Direct or Bluetooth. %In this context, it is key to determine the cooperation policy based on the content of the application.

%%% Revise
Network coding reduces the number of packet exchanges among cooperating mobile devices \cite{RandomNC}, \cite{SalimITW07}, \cite{RouayhebITW10}, \cite{SprintsonQShine10}.
%In this line of work, a block of packets are network coded and exchanged among cooperating devices until all the devices decode all the packets in the block.
Instantly decodable network coding (IDNC) considers the same problem, but focuses on instant decodability  \cite{Sadeghi_IDNC_NetCod}, \cite{Sadeghi_IDNC_Eurasip}, \cite{sorourICC}, \cite{SorourJournal}.
%\cite{IDNC}, \cite{sorourICC}, \cite{sorourGLOB}.
In particular, a network coded packet should be decodable by at least one of the devices in a cooperating group. This characteristic of IDNC makes it feasible for real-time multimedia applications in which packets are passed to the application layer immediately after they are decoded.
%if a network coded packet is decodable at a device, it is decoded and passed to the application. Otherwise, the coded packet is discarded and it is not stored for network coding opportunities available in the subsequent transmission rounds.\footnote{[][] }
Let us consider the following example to further explain the operation of IDNC.

% is a special case
%Previous work on IDNC [4]–[6], [8]–[10] focused on mini- mizing the completion delay, i.e., the time it takes to recover all the losses at all users.

\begin{figure}[t!]
\vspace{-5pt}
\centering
%\scalebox{0.5}{\includegraphics[bb=0 0 165 180]{figs_vec/example_fig_a.eps}}
\scalebox{0.38}{\includegraphics{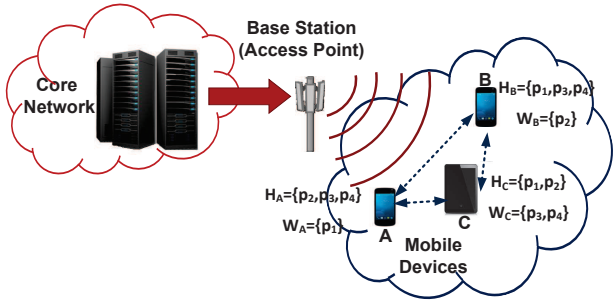}}
\vspace{-5pt}
\caption{Mobile devices $A$, $B$, and $C$ are in close proximity, and are interested in the same video content. As a simple example, let us assume that the video file is composed of four packets; $p_1$, $p_2$, $p_3$, $p_4$. Devices $A$, $B$, $C$ want to receive the sets of packets; $W_A$, $W_B$, $W_C$, respectively. They already have the sets of packets; $H_A$, $H_B$, $H_C$, respectively. %Without network coding, device $A$ transmits $p_2$ and $p_3$, and device $B$ transmits $p_1$ and $p_4$. With IDNC, device $A$ broadcasts $p_2 \oplus p_3$ and device $B$ broadcasts $p_1 \oplus p_4$. IDNC reduces the completion time from four to two transmissions.
}
\vspace{-15pt}
\label{fig:intro_example}
\end{figure}

\begin{example}
Let us consider Fig.~\ref{fig:intro_example}, where the base station broadcasts the set of the packets $\{p_1,p_2,p_3,p_4\}$ to mobile devices $A$, $B$, $C$. These devices receive the set of packets , $H_A$, $H_B$, $H_C$, successfully from the base station and want to receive the missing packets, which are the sets $W_A$, $W_B$, $W_C$, respectively. Without network coding, four transmissions are required using D2D connections, so that each device receives all the packets.
With IDNC, device $A$ broadcasts $p_2 \oplus p_3$ to devices $B$ and $C$, and device $B$ broadcasts $p_1 \oplus p_4$ to devices $A$ and $C$. After these transmissions, all devices have the complete set of packets. This example shows that IDNC has two advantages: (i) it reduces the number of transmissions from four to two, and (ii) packets are instantly decodable at each transmission; \eg when device $A$ broadcasts $p_2 \oplus p_3$, $p_2$ is decoded at device $B$ and $p_3$ is decoded at device $C$ without waiting for additional network coded packets. These advantages make IDNC feasible for real-time multimedia applications.
%
%As it is seen, IDNC reduces the completion time from four to two transmissions, so the underlying bandwidth is used more efficiently.
%Another advantage of IDNC is that its instant decodability. In this example, when device $A$ transmits $p_2 \oplus p_3$,
%One important problem, which is the focus of this paper, is the design of content-aware IDNC.
\hfill $\Box$
\end{example}

%In this work, we propose content-aware IDNC.
In the context of IDNC, the minimum amount of time that can guarantee a complete acquisition of the common content at every device is referred to as the ``completion time''. Previous works on IDNC mainly focus on reducing the completion time \cite{sorourICC}, \cite{SorourJournal}. However, the interest of each device in receiving the remaining content may vary depending on the information already received and the overall quality of service (QoS) requirements, such as bandwidth, energy, deadlines, etc. Existing network coding or IDNC schemes under such realistic constraints yield poor performance in terms of desired QoS parameters. In the following, we further illustrate on this problem.

{\em Example 1 - continued:}
Let us consider Fig.~\ref{fig:intro_example} again. Assume that there exists a constraint that devices should exchange their packets only in one transmission. (Note that IDNC requires two transmissions to deliver complete content to all devices.) This constraint may be due to (i) deadline or bandwidth; the packets may need to be played after one transmission, or (ii) energy; devices operating on batteries may put constraints on the number of transmissions. The question in this context is that which network code should be transmitted if there are such constraints, \ie a decision between $p_2 \oplus p_3$  or $p_1 \oplus p_4$ in the given transmission opportunity. This decision should be made based on the contents of the packets. The resulting optimization problem is the focus of our work in this paper.
\hfill $\Box$

We propose an efficient Content and Loss-Aware IDNC which improves content quality and network coding opportunities jointly. The following are the key contributions of this work:
\begin{itemize}
\item We consider two content-aware optimization problems: (i) {\em completion time minimization} under the quality constraint, and (ii) {\em quality maximization} under the completion time constraint.
\item We characterize the conditions that satisfy the constraints of our {\em completion time minimization} and {\em quality maximization} problems. We provide analysis of completion time and distortion by taking into account the constraints of these problems as well as the importance of each packet and the probability of channel losses over D2D links.
We develop Content-Aware IDNC algorithms for the {\em quality maximization} and {\em completion time minimization} problems based on our completion time and distortion analysis.
%We develop Content-Aware IDNC schemes for the quality maximization and completion time minimization problems using weighted IDNC graphs \cite{sorourICC}. Furthermore, we develop a novel way to determine the weights in the weighted IDNC graph by taking into account the importance of each packet, which makes our scheme content-aware.
\item We evaluate our proposed Content and Loss-Aware IDNC schemes for different number of devices and packets under the constraints of completion time and quality using real video traces. The simulation results show that Content and Loss-Aware IDNC significantly improves completion time and quality as compared to IDNC. The cost of solving the optimization problem is relatively low as we assume that the cooperation setup involves small number of devices, and each transmission phase consists of small number of packets.
\end{itemize}

The structure of the rest of the paper is as follows. Section~\ref{sec:related} presents related work. Section~\ref{sec:system} gives an overview of the system model and problem setup. Section~\ref{sec:CAIDNC} presents our Content and Loss-Aware IDNC schemes.
%the quality maximization problem under the completion time constraint. Section~\ref{sec:P2} presents the completion time minimization problem under the quality constraint.
Section~\ref{sec:perf} presents simulation results. Section~\ref{sec:conc} concludes the paper.

\section{\label{sec:related}Related Work}
Broadcasting a common content to a group of cooperating mobile devices within proximity and transmission range of each other is gaining increasing interest \cite{cisco_rep}, \cite{ericsson_rep}. In this scenario, mobile devices may receive partial content due to packet losses over wireless broadcast link. The remaining missing content can then be recovered, thanks to cooperation among the devices by exploiting D2D connections. It has been shown that random network coding \cite{RandomNC} reduces the number of transmissions necessary to satisfy all devices in the group. However, this kind of network coding, in general, requires that a block of packets be network coded and exchanged among cooperating devices until all the devices decode all packets in the block, which makes block based network coding not suitable for delay sensitive applications.

Cooperative data exchange problems have considered designing network codes to reduce the number of transmissions in the same setup.
The problem of minimizing the number of broadcast transmissions required to satisfy all devices is considered in \cite{SalimITW07}. The total number of transmissions needed to satisfy the demands of all devices, assuming cooperation among devices and the knowledge of the packet sets available in each device, is minimized in \cite{RouayhebITW10}. A deterministic algorithm that computes an optimal solution to the cooperative data exchange problem in polynomial time is proposed in \cite{SprintsonQShine10}. The cost and fairness issues of the cooperative data exchange problem have been considered in \cite{TajbakhshNetCod11}. As compared to previous cooperative data exchange problems, the focus of this paper is on instant decodability and content-awareness.

Instantly decodable network coding (IDNC) which requires instant decodability of the transmitted packets is introduced by \cite{Sadeghi_IDNC_NetCod} and \cite{Sadeghi_IDNC_Eurasip}. Minimization of the completion delay in IDNC has been considered in \cite{sorourICC}, \cite{SorourJournal}, and \cite{IDNC_cooperative}. Generalized IDNC which relaxes instant decodability constraint of IDNC to target more receivers is introduced in \cite{sorourGLOB}.
The problem of minimizing the decoding delay of generalized IDNC in persistent erasure channels is considered in \cite{SorourVTC13}.
Minimization of the broadcast completion delay for IDNC with limited feedback is considered in \cite{SorourICC11}.
Lossy feedback scenario is considered in \cite{SorourPIMRC11}.
IDNC is exploited in cooperative data exchange problem by making coding and scheduling decisions to generate IDNC packets in \cite{TajbakhshAusCTW14}. Capacity of immediately-decodable coding schemes for applications with hard deadline constraints is analyzed in \cite{CapacityOfImmDecNC}. IDNC is further relaxed in \cite{PliableNC}, where the devices are satisfied if they receive any one message that they do not have, and in \cite{AnhsRealTimeIDNC}, where the authors are interested in finding a code that is instantly decodable by the maximum number of devices. As compared to previous works on IDNC, our goal in this paper is to develop Content-Aware IDNC.

Network coding and content-awareness have met in several previous works. Multimedia video quality improvement has been considered in \cite{NguyenVideoBroadcast}, and multimedia-aware network coding scheme is developed for a broadcast and unicast scenarios for one-hop downlink topologies. One-hop opportunistic network coding scheme is considered for video streaming over wireless networks in \cite{hulya}. As compared to \cite{NguyenVideoBroadcast} and \cite{hulya}, in this paper, we consider the packet recovery problem among cooperative mobile devices using IDNC and exploiting D2D connections. Packet prioritization is considered in IDNC \cite{MuhammadICC2013}, where packet prioritization is determined based on the number of requests for a packet, whereas in this paper content-based information is used for packet prioritization.

\section{\label{sec:system}System Model \& Problem Setup}
We consider a networking model, which consists of cooperating mobile devices. Let $\Nset$ be the set of cooperating devices in our network where $N = |\Nset|$. These devices are within close proximity of each other, so they are in the same transmission range and can connect to each other via D2D links such as WiFi-Direct or Bluetooth.\footnote{Note that we do not consider any malicious or strategic activity in our setup. We rely on possible social ties in close proximity setup for cooperation incentive and to prevent any malicious or strategic behavior.}

The cooperating mobile devices in $\Nset$ are interested in receiving the packets $p_m, m=1,2,\ldots, M$ from the set $\Mset$ where $M = |\Mset|$. Packets are transmitted in two stages. In the first stage, an access point or a base station broadcasts the packets in $\Mset$ to the cooperating mobile devices in $\Nset$. In this stage, the cooperating devices may receive partial content due to packet losses over wireless broadcast link. We consider that there is no error correction mechanism in the first stage, which is dealt with in the second stage. After the first stage, the set of packets that device $n$ has is $\Hset_{n}$, and is referred to as {\em Has} set of device $n$.  The set of packets that is missing at device $n$ is, $\Lset_n$ ($\Lset_n = \Mset \setminus \Hset_n$), and is referred to as {\em Lacks} set of device $n$. Each device $n$ wants to receive all or a subset of its {\em Lacks} set, which is referred to as {\em Wants} set of device $n$ and denoted by $\Wset_{n}$.
Without loss of generality, we assume that for each packet $p_m \in \Mset$, there is at least one device that has received it successfully. In other words, $\forall p_m \in \Mset, \exists n \in \Nset \mid p_m \in \Hset_{n}$.\footnote{If there exists a packet that is lost in all devices, this packet will be sent without network coding from the base station or the access point.} We also assume that there is no packet in $\Mset$ that is received by all devices successfully. In other words, $\forall p_m \in \Mset, \exists n \in \Nset \mid p_m \in \Lset_n$.\footnote{If there exists a packet that is received successfully by all devices, we delete this packet from the set of packets, $\Mset$. Therefore, $\Mset=\cup_{n \in \Nset} \Lset_n$.}

In the second stage, the devices cooperate to recover the missing contents via their D2D connections such as WiFi-Direct or Bluetooth. Each device $n$ is satisfied after receiving the packets in its {\em Wants} set; $\Wset_{n}$. In this stage, at each transmission opportunity the best network coded packet with its corresponding transmitter is selected according to our Content and Loss-Aware IDNC algorithms which we present in the next sections. Note that network coding and cooperation decisions are made by a central device, which is selected randomly among the cooperating devices. In this setup, at each transmission opportunity, a device selected as the transmitter device by the central device, broadcasts the selected network coded packet to the other devices. The minimum amount of time that can guarantee the satisfaction of all devices $n \in \Nset$ is referred to as the ``completion time''; $T$. In this paper, $T$ is defined as the number of packet transmissions that is required for all devices to be satisfied. We denote the probability of packet loss for the D2D links by $\epsilon_{i,j}$, where $i$ is the transmitter device and $j$ is the receiver device. In particular, when the transmitter device $i$ broadcasts a packet
in the local area, device $j$ successfully receives the packet with probability $1 - \epsilon_{i,j}$. We assume that the loss probabilities $\epsilon_{i,j}, \forall i,j \in \Nset$ are i.i.d. according to a uniform distribution. The loss probabilities are predetermined by the central device as one minus the ratio of successfully received packets over transmitted packets in a time window, at the beginning of stage two.
%One of the cooperating devices is selected as the central device to make the decisions.
%If there exists a device that has all packets in the selected coded packet, it will be chosen as the sender. Otherwise, the coded packet is requested to be broadcast from the base station or access point to all devices.
%The minimum amount of time that can guarantee the satisfaction of all devices $n\in\Nset$ is referred to as the ``completion time''; $T$. In the second stage, packet loss probability over link $i-j$ is $\epsilon_{i,j}$. In particular, when a device $i$ broadcasts a packet in the local area using D2D links, device $j$ successfully receives the packet with probability $1-\epsilon_{i,j}$. We consider in our analysis that loss probability is independent and identically distributed (i.i.d) over the links and successive transmission slots. %We consider that
%transmitted packets are lost  Each device misses the broadcasted packet in the second stage according to the loss probability of the link between the device and the transmitter. We denote the loss probability of the transmitted packet from device $i$ to device $j$ by $\epsilon_{i,j}$. In other words, the broadcasted packet from device $i$ is received successfully at device $j$ with probability of $\epsilon_{i,j}$ and is lost with probability of $1-\epsilon_{i,j}$. Note that in general $\epsilon_{i,j} \neq \epsilon_{j,i}$.

In our content-aware setup, each packet $p_m \in \Mset$ has a contribution to the quality of the overall content. We refer to this contribution as the {\em importance} of packet $p_m$. The importance of packet $p_m$ for device $n$ is denoted by $r_{m,n} \geq 0$.\footnote{Note that the importance value of packets can be determined by the source and communicated to the central device so that it can make content-aware IDNC decisions. This information can be marked on a special field of the packet header. This field can be at the application level (\eg RTP headers) or part of the network coding header \cite{hulya}.}  The larger the $r_{m,n}$, the more important packet $p_m$ is for device $n$. For example, in applications that the content is video or image, $r_{m,n}$ is calculated as the distortion of the content that device $n$ experiences from lacking packet $p_m$. Therefore, the distortion value for device $n$ is calculated as:
\begin{equation} \label{eq:distortion}
 D_n=\sum_{m | p_m \in \Mset} r_{m,n} - \sum_{m | p_m \in \Hset_n} r_{m,n}.
\end{equation}

The goal of traditional IDNC is ``to minimize $T$'' \cite{sorourICC}, \cite{SorourJournal}. On the other hand, Content and Loss-Aware IDNC takes into account packet importances and distortion value $D_n$ formulated in Eq.~(\ref{eq:distortion}). In addition, we take into account the packet losses of D2D links in our formulations. In particular, we consider the following two problems:
\begin{itemize}
\item \underline{Content and Loss-Aware IDNC-P$_1$}: Our first problem minimizes the completion time $T$ under the quality constraint.
\begin{align}
\label{eq:CAIDNCP1-O} \mbox{minimize    } & T\\
\label{eq:CAIDNCP1-C} \mbox{subject to  } & D_n \leq D^{cons}_n, \mbox{  } \forall n \in \Nset
\end{align} where $D^{cons}_n$ is the maximum tolerable distortion for device $n$. This problem is relevant if there are limitations on the number of transmissions due to available bandwidth or energy. \Eg if devices are conservative in terms of their energy consumption, then the correct problem is to minimize the number of transmissions, which is equivalent to minimizing the completion time $T$, while satisfying a quality constraint; Eq.~(\ref{eq:CAIDNCP1-C}).
\item \underline{Content and Loss-Aware IDNC-P$_2$}: Our second problem maximizes quality under the completion time constraint.
\begin{align}
\label{eq:CAIDNCP2-O} \mbox{minimize    } & f(\textbf{D})\\
\label{eq:CAIDNCP2-C} \mbox{subject to  } & T \leq T^{cons},
\end{align} where $T^{cons}$ is the maximum allowed completion time, $\textbf D$ is the vector of per device distortions; $\textbf{D} = [D_1,D_2,...,D_N]$, and $f(\textbf{D})$ is the function of the distortion vector; $\textbf{D}$. $f(\textbf{D})$ should be a convex function, and depending on the application, it can take different values \cite{distortion_function}. For example, in some applications the goal may be to  minimize the sum distortion over all devices; \ie $f(\textbf{D})=\sum_{n\in\Nset} D_n$, while in some other applications the goal may be to minimize the maximum distortion over all devices; $f(\textbf{D})=\max_{n\in\Nset} D_n$ \cite{distortion_function}. We further explain our approach to select $f(\textbf{D})$ in Section~\ref{sec:CAIDNC}. The problem of minimizing $f(\textbf{D})$ is relevant if there are constraints on delay. \Eg if packets should be played out before a hard-deadline constraint; $T^{cons}$, then the goal is to improve the content quality as much as possible; Eq.~(\ref{eq:CAIDNCP2-O}), before the deadline; Eq.~(\ref{eq:CAIDNCP2-C}).
\end{itemize}

In the next section, we provide our solutions to {\em Content and Loss-Aware IDNC-P$_1$} and {\em Content and Loss-Aware IDNC-P$_2$}. 

\section{\label{sec:CAIDNC} Content and Loss-Aware IDNC}
\subsection{Minimizing Completion Time under Quality Constraint}
In this section, we present our approach to solve the problem; Content and Loss-Aware IDNC-P$_1$ in Eqs.~(\ref{eq:CAIDNCP1-O}), (\ref{eq:CAIDNCP1-C}).

{\bf Strategy to solve the problem:} The main challenge while solving the optimization problem in Eqs.~(\ref{eq:CAIDNCP1-O}), (\ref{eq:CAIDNCP1-C}) comes from the fact that the closed form expression for the completion time; $T$ is an open problem. One possible approach, as also considered in previous work \cite{sorourICC}, \cite{NguyenVideoBroadcast}, is to formulate the problem as a Markov decision process, and we consider a similar approach in this paper as explained next.

Let {\em state $s$} be the set of {\em Has} sets of all devices, {\em action $a$} be the selection and transmission of an IDNC packet, and the {\em terminating state} is any state, for which the constraint on the distortion values; Eq.~(\ref{eq:CAIDNCP1-C}) is satisfied. By considering the packet losses of D2D links, the system moves to one of the states in the set $\Sset_{a,s}^*$ from state $s$, by taking an action $a$. We define the completion time $T$ and $T^*$ as the number of packets, required to be transmitted and received successfully at the targeted receivers, to reach the termination state (any state for which Eq.~(\ref{eq:CAIDNCP1-C}) is satisfied) from state $s$ and $s^* \in \Sset_{a,s}^*$, respectively.
%Also, $T^*$ is defined as the average of completion time over all states in the set $\Sset_{a,s}^*$.
Our approach is to take the action that results in the minimum average of completion time over all next states in the set $\Sset_{a,s}^*$.
%, where $T^*$ is defined as the number of packets required to be transmitted to reach the termination state (any state for which Eq.~(\ref{eq:CAIDNCP1-C}) is satisfied) from state $s^*$.
Motivated by this fact, we next estimate the completion time at the current state; $T$ as well as the next state; $T^*$.

{\bf Relating Completion Time to ``{\em Wants} Sets'':}
In our setup, as different from previous work \cite{sorourICC}, each device does not have a fixed initial {\em Wants} set. Instead, each device is interested in receiving any set of packets so that Eq.~(\ref{eq:CAIDNCP1-C}) is satisfied. Indeed, for device $n$, $L_n$ different {\em Wants} sets; $\Wset_n^l, l=1,2,\dots,L_n$ could satisfy Eq.~(\ref{eq:CAIDNCP1-C}) as long as the following conditions are met.

\begin{itemize}
\item
C$_1$: $\Wset_n^l \subseteq \Lset_n.$\ \\
\item
C$_2$: $\mathlarger{\mathlarger{‎‎\sum}}_{m | p_m \in \Mset} r_{m,n}-\mathlarger{\mathlarger{‎‎\sum}}_{m | p_m \in (\Wset_n^l \cup \Hset_n)} r_{m,n} \leq D^{cons}_n.$\ \\
%\begin{equation} \nonumber
%$ \mathlarger{\mathlarger{‎‎\sum}}_{m | p_m \in \Mset} r_{m,n}-\mathlarger{\mathlarger{‎‎\sum}}_{m | p_m \in (\mathcal{W}_n^l \cup \mathcal{H}_n)} r_{m,n} \leq $D^{cons}_n.
%\end{equation}
\item
C$_3$: If $\Wset_n^{l_1} \supset \Wset_n^{l_2}, l_1,l_2=1,2,\dots,L_n$ then delete $\Wset_n^{l_1}$.\ \\
\end{itemize}
It is obvious that a {\em Wants} set should be a subset of the {\em Lacks} set (the first condition; C$_1$). The second condition; C$_2$ is required to satisfy the constraint of our problem, \ie Eq.~(\ref{eq:CAIDNCP1-C}). The third condition; C$_3$ picks the set with the minimum cardinality between each pair of sets that are superset/subset of each other and deletes the other one. This condition is required to reach our objective of minimizing the number of packets to be transmitted. Let us explain the conditions; C$_1$, C$_2$, C$_3$ via the following example.

\begin{example}
Assume that device $n \in \Nset$ is interested in receiving $M=4$ packets with the importance values of: $r_{1,n}=4, r_{2,n}=5, r_{3,n}=3, r_{4,n}=1$, device $n$'s {\em Has} set is $\Hset_n=\left\{p_1\right\}$, and its maximum tolerable distortion is equal to $D^{cons}_n=5$. By applying the first and the second conditions, the potential {\em Wants} sets are: $\Wset_n^1=\left\{p_2\right\}, \Wset_n^2=\left\{p_3,p_4\right\}, \Wset_n^3=\left\{p_2,p_4\right\}, \Wset_n^4=\left\{p_2,p_3\right\}, \Wset_n^5=\left\{p_2,p_3,p_4\right\}$. According to the third condition, ($\Wset_n^3 \supset \Wset_n^1, \Wset_n^4 \supset \Wset_n^1, \Wset_n^5 \supset \Wset_n^1$), only the first two sets are kept as potential {\em Wants} sets: $\Wset_n^1=\left\{p_2\right\}, \Wset_n^2=\left\{p_3,p_4\right\}$.
\hfill $\Box$
\end{example}

Now that we defined {\em Wants} sets for our problem, we can formulate the completion time in terms of {\em Wants} sets as follows. The completion time for device $n$, denoted by $T_n$, is equal to the minimum number of packets that it should receive successfully so that its distortion is equal to or less than its maximum tolerable distortion:
\begin{align}
T_n=\min_{l=1,\dots,L_n} |\Wset_n^{l}|.
\label{eq:Tn}
\end{align} Note that device $n$ can benefit from a transmitted IDNC packet, if it is instantly decodable for device $n$ and the decoded packet is a member of the set $\bigcup_{l=1}^{L_n} \Wset_n^l$. Assume that device $n$ receives the transmitted packet successfully and decodes packet $p_m$, then the system moves from state $s$ to state $s^* \in \Sset_{a,s}^*$. The completion time and the potential {\em Wants} sets for device $n$ at state $s^*$ are expressed as:
\begin{align} \label{eq:CT}
    T^*_n=
\begin{cases}
    T_n-1,& \text{if } \exists l \leq L_n \mid (p_m \in \Wset_n^{l} \quad\\
    & \quad \quad \quad \quad \quad \& |\Wset_n^{l}|=T_n)\\
    T_n,              & \text{otherwise}
\end{cases}
\end{align}
\begin{align}
(\Wset_n^{l})^* = (\Wset_n^{l} \setminus p_m), l=1,2,\ldots,L_n.
\end{align}

{\bf Lower and Upper Bounds of $T$:}
%\begin{theorem}\label{theorem:bounds_CT}
The completion time, $T$, which is the minimum number of packets required to be transmitted and received successfully at the targeted receivers to reach the terminating state, has the lower and upper bounds of:
\begin{align} \label{eq:LowUpBound}
\max_{n \in \Nset} T_n \leq T \leq \sum_{n \in \Nset} T_n.
\end{align}
%\end{theorem}
%{\em Proof:}
In particular, each device $n$ needs at least $T_n$ packet transmissions to be satisfied. In the worst case scenario, at each transmission, only one of the devices is targeted and its completion time is reduced by one if it receives the transmitted packet successfully. Therefore, $\sum_{n \in \Nset} T_n$ transmissions are required. This is equal to the upper bound of completion time; $T \leq \sum_{n \in \Nset} T_n$. On the other hand, the device with the maximum completion time needs to receive $\max_{n \in \Nset} T_n$ transmissions successfully. In the best case scenario, the $\max_{n \in \Nset} T_n$ packet transmissions can be chosen so that the other devices can also be targeted and satisfied by these transmissions. Note that a single transmission can benefit a subset of devices if they want the same packet or if the wanted packets are network coded in the single transmission. The bounds in Eq.~(\ref{eq:LowUpBound}) are explained via the next example.

\begin{example}
Let us consider three devices with the completion times of $T_1=1, T_2=2, T_3=3$. Obviously, device 3 needs at least three transmissions, $T_3=3$, to be satisfied, \ie to receive all the packets in its {\em Wants} set with the minimum size, $\min_{l=1,\dots,L_3} |\Wset_3^{l}|$ (Eq.~(\ref{eq:Tn})). In the best case scenario, these three transmissions can also target and satisfy the other two devices. In other words, according to Eq.~(\ref{eq:CT}), the completion time is decreased by one for device 3 in all three transmissions, the completion time is decreased by one for device 2 in two of the transmissions and the completion time is decreased by one for device 1 in just one of the transmissions. Therefore, the lower bound for the completion time is (Eq.~(\ref{eq:LowUpBound})) $T=\max_{n \in \Nset} T_n=3$. On the other hand, in the worst case scenario, at each successful transmission, just one of the devices is targeted and satisfied. Therefore, 3 transmissions are required to be received successfully at device 3 and satisfy it, 2 transmissions are required to be received successfully at device 2 and satisfy it, and 1 transmission is required to be received successfully at device 1 and satisfy it. Therefore, the upper bound for the completion time is $T = \sum_{n \in \Nset} T_n=1+2+3=6$. In general, the completion time varies between the lower and upper bounds in Eq.~(\ref{eq:LowUpBound}).
\hfill $\Box$
\end{example}

{\bf Expressing $T^*$ as a $p-$norm:}
As we mentioned earlier, our approach to solve Content and Loss-Aware IDNC-P$_1$ is to take the action, \ie selecting the network code, that results in the next states $s^* \in \Sset_{a,s}^*$ with the minimum average completion time. Consider the next state $s^* \in \Sset_{a,s}^*$ with the completion time $T^*$. Although we do not have analytically closed form formulation for the completion time; $T$, hence $T^*$, we have lower and upper bounds on $T$; Eq.~(\ref{eq:LowUpBound}), which also applies to $T^*$:
\begin{align} \label{eq:LowUpBoundT'}
\max_{n \in \Nset} T^*_n \leq T^* \leq \sum_{n \in \Nset} T^*_n,
\end{align} where $T^*_n$ is characterized by Eq.~(\ref{eq:CT}).

Our goal is to find the network code that minimizes the average of $T^*$ among all next states $s^* \in \Sset_{a,s}^*$, so let us examine the lower and upper bounds of $T^*$ closely. The lower bound of $T^*$ is $\max_{n \in \Nset} T^*_n$ which is actually the maximum norm (infinity norm or $\textit{L}_\infty$ norm) of the vector $\textbf{\textit{T}}^*=[T^*_1,T^*_2,...,T^*_N]$, \ie the maximum norm of $\textbf{\textit{T}}^*$ is expressed as $\left\|\textbf{\textit{T}}^*\right\|_\infty = \max_{n \in \Nset} T^*_n$. On the other hand, the upper bound of $T^*$ is $\sum_{n \in \Nset} T^*_n$, which is the $\textit{L}_1$ norm of the vector $\textbf{\textit{T}}^*=[T^*_1,T^*_2,...,T^*_N]$, \ie the $\textit{L}_1$ norm of $\textbf{\textit{T}}^*$ is expressed as $\left\|\textbf{\textit{T}}^*\right\|_1 = \sum_{n \in \Nset} T^*_n$. Thus, $\left\|\textbf{\textit{T}}^*\right\|_\infty \leq T^* \leq \left\|\textbf{\textit{T}}^*\right\|_1$. Since the following inequality holds; $\left\|\textbf{\textit{T}}^*\right\|_\infty \leq \left\|\textbf{\textit{T}}^*\right\|_p \leq \left\|\textbf{\textit{T}}^*\right\|_1$, we can conclude that $T^* = \left\|\textbf{\textit{T}}^*\right\|_p$ for some $p$ such that $1 < p < \infty$. Now that we know $T^* = \left\|\textbf{\textit{T}}^*\right\|_p$, we can select a network code which minimizes the p-norm of the average completion time among all next states; $\left\|\bar{\textbf{\textit{T}}^*}\right\|_p$.\footnote{Note that by minimizing $\left\|\bar{\textbf{\textit{T}}^*}\right\|_p$, instead of minimizing $\bar{T^*}$ itself, we loose optimality as we do not know the exact value of $p$. However, this relaxation allows us to tackle the problem. Furthermore, simulation results show that this approach provides significant improvement. The performance of IDNC for various $p$ values is analyzed in \cite{SorourJournal}. In this paper, we consider $p=2$.}

{\bf Taking Action:}
Since our goal is to select a network code which minimizes $\left\|\bar{\textbf{\textit{T}}^*}\right\|_p$, we should determine all possible instantly decodable network coding candidates. Then, we should select the best network code which minimizes $\left\|\bar{\textbf{\textit{T}}^*}\right\|_p$. A trivial approach would be exhaustively listing all possible network coding candidates, and calculating $\left\|\bar{\textbf{\textit{T}}^*}\right\|_p$ for each of them to determine the best one. More efficient approach is to use a graph; IDNC graph \cite{sorourICC}, \cite{SorourJournal}. IDNC graph is constructed so that each clique in the graph corresponds to a network code. Thus, we can find the best clique to determine the best network code which minimizes $\left\|\bar{\textbf{\textit{T}}^*}\right\|_p$.

The IDNC graph $\mathcal{G}$ for our cooperative data exchange system consists of $N$ disjoint IDNC local graphs. Each IDNC local graph $\mathcal{G}_t, t \in \Nset$ represents the network coding candidates that can be transmitted from device $t$.
The IDNC local graph $\mathcal{G}_t$ for our problem is constructed as follows.

For device $n \in (\Nset \setminus t)$, $|(\bigcup_{l=1,\ldots,L_n} \mathcal{W}_{n}^l) \cap \Hset_t|$ vertices, each shown by $v_{n,m}^t$ such that $p_m\in(\bigcup_{l=1,\ldots,L_{n}} \mathcal{W}_{n}^l) \quad \& \quad p_m \in \mathcal{H}_t$ are added to the graph. A pair of vertices, $v_{n,m}^t$ and $v_{k,l}^t$, are connected if one of the following conditions; C$_{1}^{'}$ or C$_{2}^{'}$ is satisfied:
\begin{itemize} \label{eq:connect}
\item C$_{1}^{'}$: $p_m=p_l$
\item C$_{2}^{'}$: $p_m \in \mathcal{H}_k \quad \& \quad p_l \in \mathcal{H}_n$.
\end{itemize}

The total number of possible actions when device $t$ is the transmitter, \ie the number of network codes that device $t$ can transmit, is equal to the number of cliques in the local graph $\mathcal{G}_t$. The action associated with clique $q_t \in \mathcal{G}_t$ corresponds to transmitting the network coded packet generated by XORing all the packets associated with the clique, \ie XORing $\forall p_m$ such that $v_{n,m}^t \in q_t$. The best network code that can be transmitted from device $t$, hence the best clique in $\mathcal{G}_t$ is the one that minimizes $\left\|\bar{\textbf{\textit{T}}^*}\right\|_p$. We assign weights to each vertex in the graph so that the sum weight of all the vertices in clique $q_t$ corresponds to $\left\|\bar{\textbf{\textit{T}}^*}\right\|_p$ which is resulted from sending the network code represented by the clique $q_t$ from device $t$.
Then, we consider graph $\mathcal{G}$, which is equal to the union of all local graphs; $\mathcal{G}=\bigcup_{t \in \Nset} \mathcal{G}_t$ and search for the clique that has the largest total weight summed over its vertices. Next, we determine the weight of vertex $v_{n,m}^t$ in clique $q_t$; $w_{n,m}^t$.
%The weights assigned to each clique in the IDNC graph is in the direction to reach our ultimate goal from the SSP problem; selecting the action that results in a state with the minimum completion time. We explain the details next.

Assume that the network code corresponding to clique $q_t$ is transmitted from device $t$. This packet is received successfully by any device $n$ with probability of $1-\epsilon_{t,n}$ and is lost with probability of $\epsilon_{t,n}$. Therefore, the average of the resulting completion times, $\bar{T^*}$, will have $p-$norm; $\left\|\bar{\textbf{\textit{T}}^*}\right\|_p$, which is equal to:
%Assume that clique $q$ with $L_q$ vertices in our IDNC graph $\mathcal{G}$ is selected. By sending the packet correspondent to clique $q$, the $\textit{L}_p$ norm distance of the resulting point from the origin will be:
\begin{align} \label{eq:distance}
 \left\|\bar{\textbf{\textit{T}}^*}\right\|_p = (&\sum_{n|(\exists p_m | v_{n,m}^t \in q_t)} ((1-\epsilon_{t,n})T^*_n+\epsilon_{t,n}T_n)^p \nonumber \\
 &+\sum_{n|( \nexists p_m | v_{n,m}^t \in q_t)} (T_n)^p)^{1/p}
\end{align}
Note that the completion time for device $n$ changes from $T_n$ to $T^*_n$ (Eq.~(\ref{eq:CT})) with probability of $(1-\epsilon_{t,n})$ and does not change with probability of $\epsilon_{t,n}$ if the selected clique covers device $n$, \ie it includes a vertex that represents a packet from Wants set of device $n$. The term $\sum_{n|(\exists p_m | v_{n,m}^t \in q_t)} ((1-\epsilon_{t,n})T^*_n+\epsilon_{t,n}T_n)^p$ in Eq.~(\ref{eq:distance}) corresponds to this fact. On the other hand, the completion time for the devices that are not covered by the selected clique does not change. The term $\sum_{n|( \nexists p_m | v_{n,m}^t \in q_t)} (T_n)^p$ in Eq.~(\ref{eq:distance}) corresponds to this fact. Eq.~(\ref{eq:distance}) is expressed as;
\begin{align} \label{eq:distance_v2}
&\left\|\bar{\textbf{\textit{T}}^*}\right\|_p = (\sum_{n \in \Nset} (T_n)^p + \nonumber \\
&\sum_{n|(\exists p_m | v_{n,m}^t \in q_t)} ((1-\epsilon_{t,n})T^*_n+\epsilon_{t,n}T_n)^p-(T_n)^p)^{1/p}
\end{align}
Note that the first term in Eq.~(\ref{eq:distance_v2}) is the same and fixed for all cliques in the graph. Therefore, in order to minimize $\left\|\bar{\textbf{\textit{T}}^*}\right\|_p$, the second term should be minimized, which corresponds to:
\begin{align}
\label{eq:q}
q_t^*=\operatorname*{arg\,max}_{q_t} &\sum_{n|(\exists p_m | v_{n,m}^t \in q_t)} ((T_n)^p- \nonumber \\
& \quad \quad \quad \quad ((1-\epsilon_{t,n})T^*_n+\epsilon_{t,n}T_n)^p).
\end{align}
where $q_t^*$ is the best clique and the corresponding network code is the best network code in the local graph $\mathcal{G}_t$. By substituting $T^*_n$ from Eq.~(\ref{eq:CT}) into Eq.~(\ref{eq:q}), the following weight assignment to vertex $v_{n,m}^t \in \mathcal{G}_t$ is obtained:
\begin{align} \label{eq:weights}
    w_{n,m}^t=
\begin{cases}
    (T_n)^p-(T_n-1+\epsilon_{t,n})^p,& \text{if } \exists l \leq L_n \mid\\
    & p_m\in\mathcal{W}_n^{l}\\
		& \& |\mathcal{W}_n^{l}|=T_n\\
    0,              & \text{otherwise.}
    \end{cases}
\end{align}
Using the weight assignments in Eq.~(\ref{eq:weights}), Content and Loss-Aware IDNC-P$_1$ finds the network code that corresponds to the maximum weighted clique in graph $\mathcal{G}$ at each transmission opportunity until Eq.~(\ref{eq:CAIDNCP1-C}) is satisfied. Note that $\mathcal{G}$ is the union of all local graphs $\mathcal{G}_t$, $\mathcal{G}=\bigcup_{t \in \Nset} \mathcal{G}_t$.

\subsection{Maximizing Quality under Completion Time Constraint}
In this section, we present our approach to solve the problem; Content and Loss-Aware IDNC-P$_2$ presented in Eqs.~(\ref{eq:CAIDNCP2-O}),
(\ref{eq:CAIDNCP2-C}).
For the solution of Content and Loss-Aware IDNC-P$_2$, we use a similar approach to the solution of Content and Loss-Aware IDNC-P$_1$.

{\bf Expressing $f(\textbf{D})$ as a $p-$norm:}
As we mentioned earlier, depending on the application, the distortion function $f(\textbf{D})$ can take different values \cite{distortion_function}. If the goal is to  minimize the sum distortion over all devices, then $f(\textbf{D}) = \sum_{n\in\Nset} D_n$ which is actually the  $\textit{L}_1$ norm of the distortion vector; $\textbf{D} = [D_1,D_2,...,D_N]$, \ie $\left\|\textbf{\textit{D}}\right\|_1 = \sum_{n \in \Nset} D_n$. On the other hand, if the goal is to minimize the maximum distortion over all devices, then $f(\textbf{D}) = \max_{n\in\Nset} D_n$, which is actually the maximum (infinity) norm of the distortion vector; $\textbf{D} = [D_1,D_2,...,D_N]$, \ie $\left\|\textbf{\textit{D}}\right\|_\infty = \max_{n \in \Nset} D_n$. For the sake of generality, we consider the objective function as $p-$norm of the distortion vector; $f(\textbf{D})=\left\|\textbf{\textit{D}}\right\|_p$, $\forall p \geq 1$.

{\bf Taking Action:}
Since our goal is to select a network code which minimizes $f(\textbf{D})=\left\|\textbf{\textit{D}}\right\|_p$, we should determine all possible instantly decodable network coding candidates. Then, we should select the best network code which minimizes $\left\|\textbf{\textit{D}}\right\|_p$. As we discussed in the solution of Content and Loss-Aware IDNC-P$_1$, a trivial approach would be exhaustively listing all possible network coding candidates, and calculating $\left\|\textbf{\textit{D}}\right\|_p$ for each of them to determine the best one. However, constructing IDNC graph is more efficient. In Content and Loss-Aware IDNC-P$_2$, the local IDNC graph $\mathcal{G}_t$ is constructed as follows. For device $n$, $|\mathcal{L}_n \cap \Hset_t|$ vertices, each shown by $v_{n,m}^t$ such that $p_m\in \mathcal{L}_n \quad \& \quad p_m \in \mathcal{H}_t$ are added to the graph. The vertex $v_{n,m}^t$ represents the missing packet $p_m$ (with priority of $r_{m,n}$) in device $n$ that can be transmitted from device $t$. The vertices in the graph are connected according to the rules C$_{1}^{'}$ and C$_{2}^{'}$ presented in the previous section. Each clique in the graph represents a network coded packet. Assume that the network code corresponding to clique $q_t \in \mathcal{G}_t$ is transmitted from device $t$. This packet is received successfully by any device $n$ with probability of $1-\epsilon_{t,n}$ and is lost with probability of $\epsilon_{t,n}$. Therefore, $p-$norm of the distortion is equal to:
\begin{align} \label{eq:distance_p2}
 \left\|\textbf{\textit{D}}\right\|_p & = (\sum_{n|(\exists p_m |v_{n,m}^t \in {q_t})} (\epsilon_{t,n}(D_n) \nonumber \\
 &+(1-\epsilon_{t,n})(D_n-r_{m,n}))^p \nonumber \\
 &+ \sum_{n|(\nexists p_m |v_{n,m}^t \in {q_t})} (D_n)^p)^{1/p} \nonumber \\
 & = (\sum_{n\in\Nset} (D_n)^p+ \nonumber \\
 & \sum_{n|(\exists p_m | v_{n,m}^t \in q_t)} (D_n-r_{m,n}+\epsilon_{t,n}r_{m,n})^p \nonumber \\
 & -(D_n)^p)^{1/p}.
\end{align}
Note that the first term in the above equation, $\sum_{n\in\Nset} (D_n)^p$, is the same and fixed for all cliques in the graph. In order to minimize $\left\|\textbf{\textit{D}}\right\|_p$ in Eq.~(\ref{eq:distance_p2}), the second term should be minimized. Therefore, the weight assigned to vertex $v_{n,m}^t \in \mathcal{G}_t$ is equal to:
\begin{align}
w_{n,m}^t=(D_n)^p-(D_n-r_{m,n}+\epsilon_{t,n}r_{m,n})^p.
\end{align}
Our algorithm Content and Loss-Aware IDNC-P$_2$ selects the clique with the maximum weight summed over its vertices in the graph $\mathcal{G}=\bigcup_{t \in \Nset}\mathcal{G}_t$. A network code corresponding to the maximum weighted clique is selected and transmitted to all devices from the corresponding transmitter.

\section{\label{sec:perf}Simulation Results}
We implemented the proposed Content and Loss-Aware IDNC schemes by considering that there may be losses over D2D connections, and compared it with three baselines:

\begin{itemize}
\item Content-Aware Loss-Unaware IDNC: This scheme is proposed in our previous work \cite{ICNC} assuming that D2D connections are lossless. In this method, each vertex in the local graph has a weight which is based on its contribution to minimizing the completion time for Content-Aware IDNC-P$_1$ and minimizing the distortion function for Content-Aware IDNC-P$_2$ without considering the probability of successful reception of the transmitted packet. As different from \cite{ICNC}, in this paper the network coded packets that have higher probability of successful reception as well as smaller completion time for Content-Aware IDNC-P$_1$ and smaller distortion function values for Content-Aware IDNC-P$_2$ are selected at each transmission opportunity. Our proposed method in this paper outperforms the method in \cite{ICNC} when D2D connections are lossy, as shown in the simulation results.

\item Loss-Aware IDNC: This scheme, proposed in \cite{IDNC_cooperative}, takes into account the probability of D2D link losses among the cooperative devices, but it is not content-aware. In this method, the weight assignment to each vertex in the local graph is based on the sizes of its targeted  receivers' \emph{Lacks} sets as well as the successful reception of the transmitted packet at the targeted receivers. Our proposed method in this paper outperforms the method in \cite{IDNC_cooperative} under the realistic constraints of delay and quality, as shown in the simulation results.

\item Loss-Unaware IDNC: This scheme does not take into account the probability of channel losses over D2D connections among the cooperative devices. In addition, it is not content-aware. In \cite{SorourJournal}, an IDNC scheme is proposed for recovering the missing content through broadcasting IDNC packets from the base station or the access point. We use an adopted version of this method for our cooperative system setup, called Loss-Unaware IDNC, for comparison. We consider local IDNC graphs; then according to \cite{SorourJournal}, each vertex in the local graph is assigned a weight which is based on sizes of the \emph{Lacks} sets for its targeted receivers as well as sizes of the \emph{Lacks} sets for the devices targeted by its adjacent's vertices. Our proposed method in this paper outperforms Loss-Unaware IDNC scheme when there are losses over D2D links and under the realistic constraints of delay and quality, as shown in the simulation results.

\end{itemize}

\begin{figure}[t!]
%\vspace{-5pt}
\begin{center}
\subfigure[Completion Time vs. Number of Devices]{\includegraphics[scale=0.45]{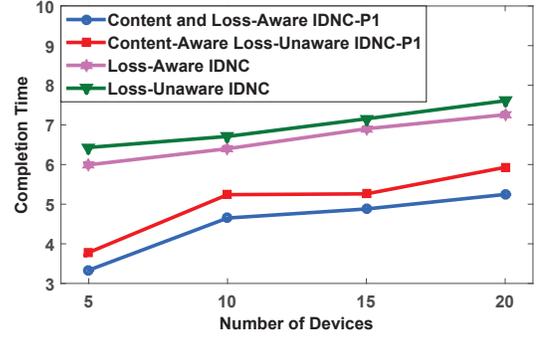}}
\subfigure[Completion Time vs. Number of Packets]{\includegraphics[scale=0.45]{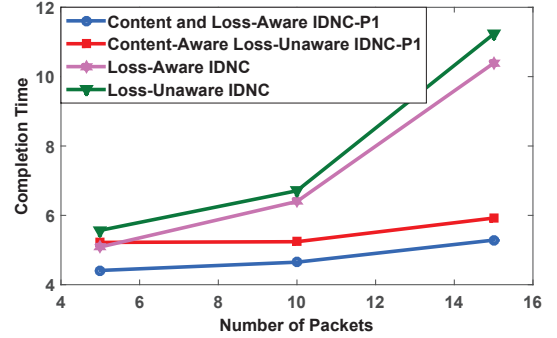}}
\subfigure[Completion Time vs. Constrained Distortion]{\includegraphics[scale=0.45]{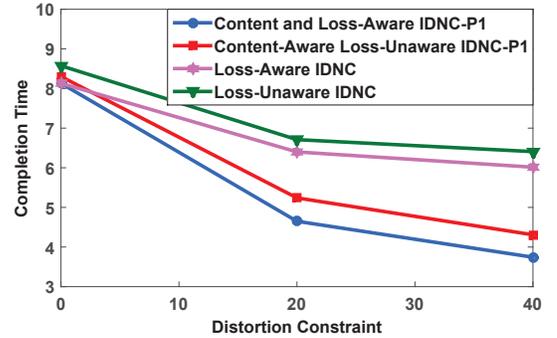}}
\end{center}
\begin{center}
\vspace{-10pt}
\caption{\label{performance_figs_CT} The performance of Content and Loss-Aware IDNC-P$_1$, Content-Aware Loss-Unaware IDNC-P$_1$, Loss-Aware IDNC, and Loss-Unaware IDNC.}
\vspace{-10pt}
\end{center}
\end{figure}

\begin{figure}[t!]
%\vspace{-5pt}
\begin{center}
\subfigure[$f(\textbf{D})=\sqrt{\sum_{n\in Nset} D_n^2}$ vs. Number of Devices]{\includegraphics[scale=0.45]{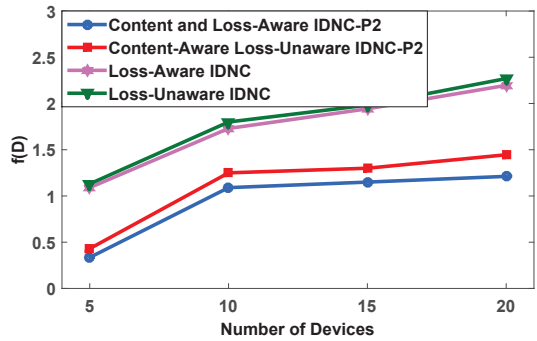}}
\subfigure[$f(\textbf{D})=\sqrt{\sum_{n\in Nset} D_n^2}$ vs. Number of Packets]{\includegraphics[scale=0.45]{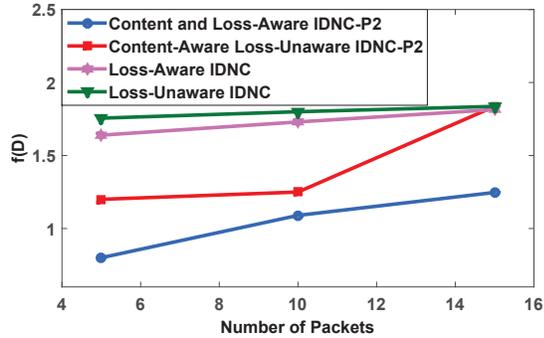}}
\subfigure[$f(\textbf{D})=\sqrt{\sum_{n\in Nset} D_n^2}$ vs. Constrained Completion Time]{\includegraphics[scale=0.45]{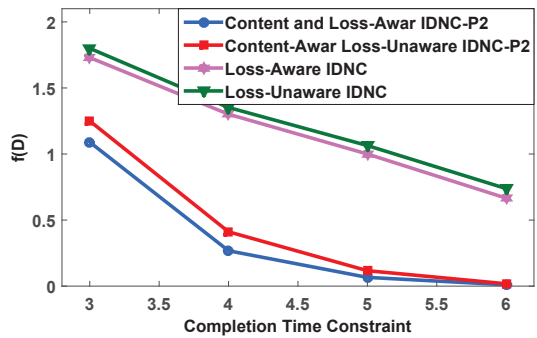}}
\end{center}
\begin{center}
\vspace{-10pt}
\caption{\label{performance_figs_D} The performance of Content and Loss-Aware IDNC-P$_2$, Content-Aware Loss-Unaware IDNC-P$_2$, Loss-Aware IDNC, and Loss-Unaware IDNC.}
\vspace{-10pt}
\end{center}
\end{figure}
%\begin{figure}[!t]
%\centering
%\includegraphics[scale=0.4]{Fig/CT_D_u5p10.png}
%\caption{Completion Time vs. Constrained Distortion}
% \label{fig:CTdistortion}
%\end{figure}

We consider a topology shown in Fig.~\ref{fig:intro_example} for different number of devices. First all packets are broadcast from the source in phase 1. Each device selects its loss probability uniformly from the region $[0.3,0.8]$ for Figs. \ref{performance_figs_CT} and \ref{performance_figs_D}, and misses packets according to the selected loss probability. Then, in phase 2, the devices cooperate to recover the missing packets. The probability of loss for a packet transmission from device $i$ to device $j$, $\epsilon_{i,j}, i\in\Nset, j\in\Nset$ is selected from a uniform distribution in the region $[0,0.3]$ for Figs. \ref{performance_figs_CT} and \ref{performance_figs_D}.
%In the graphs the completion time is defined as the number of packet transmissions and the distortion is defined according to Eq.~(\ref{eq:distortion}).

{\bf Completion Time \& Distortion:}
Fig.~\ref{performance_figs_CT}(a) and \ref{performance_figs_CT}(b) show the completion time required by Content and Loss-Aware IDNC-P$_1$, Content-Aware Loss-Unaware IDNC-P$_1$, Loss-Aware IDNC, and Loss-Unaware IDNC under the constraint of $D^{cons}_n=0.2 \sum_{m | p_m \in \Mset} r_{m,n}$ for device $n$. In this setup, $r_{m,n}$ is generated according to a gamma distribution with mean 1 and variance 50. Fig.~\ref{performance_figs_CT}(a) shows the results for transmitting 10 packets to different number of devices. Fig.~\ref{performance_figs_CT}(b) shows the results for transmitting different number of packets to 10 devices. In these graphs, the required completion time increases with increasing number of devices/packets. As seen, the completion time using Content and Loss-Aware IDNC-P$_1$, is smaller than the other methods. %As expected, the required completion time is the largest when no No-NC is used.

Fig.~\ref{performance_figs_CT}(c) shows the required completion time for sending 10 packets to 10 devices, under the constraint of 0, 20\%, and 40\% distortion for each device. As expected, under the constraint of no distortion (\ie all packets are demanded by all devices), the performance of Content and Loss-Aware IDNC-P$_1$ and Loss-Aware IDNC are almost the same and better than Content-Aware Loss-Unaware IDNC-P$_1$ and Loss-Unaware IDNC. The more the tolerable distortion, the more improvement is observed by Content and Loss-Aware IDNC-P$_1$.

Fig.~\ref{performance_figs_D}(a) and \ref{performance_figs_D}(b) show the distortion function of $f(\textbf{D})=\sqrt{\sum_{n\in Nset} D_n^2}$ for Content and Loss-Aware IDNC-P$_2$, Content-Aware Loss-Unaware IDNC-P$_2$, Loss-Aware IDNC, and Loss-Unaware IDNC under the constraint that $T^{cons}=3$. Fig.~\ref{performance_figs_D}(a) shows the results for transmitting 10 packets to different number of devices and Fig.~\ref{performance_figs_D}(b) shows the results for transmitting different number of packets to 10 devices. As shown in the figures, the performance is improved significantly using Content and Loss-Aware IDNC-P$_2$. %Distortion value using No-NC is the largest among all.

{\bf Real Video Traces:} Table~\ref{table:VT} shows the results for the total distortion improvement of Content and Loss-Aware IDNC-P$_2$ over Content-Aware Loss-Unaware IDNC-P$_2$ , Loss-Aware IDNC and Loss-Unaware IDNC for real video traces (Akiyo and Grandma) under completion time constraint. Our video traces are CIF sequences encoded using the JM 8.6 version of the H.264/AVC codec \cite{h264_1}, \cite{h264_2}. Each video trace is divided into blocks of packets, where block size is 10. The importance of each packet is determined by its contribution to overall video quality. The importance of each packet was determined by removing it from the video sequence, and measuring the total video quality distortion (when the packet is missing) using our H.264/AVC video codec.
The video packets are delivered to 10 devices. Each device selects its loss probability in phase 1 uniformly from the region $[0.3 0.4]$ and the loss probability for the D2D link between each two devices in phase 2 is selected uniformly from the region $[0 0.5]$.%The percentage of total distortion improvement of Content-Aware IDNC-P$_2$  as compared to IDNC and No-NC is shown in the table.

As seen, Content and Loss-Aware IDNC-P$_2$ improves by $14.5\%$ over Content-Aware Loss-Unaware IDNC, $7.3\%$ over Loss-Aware IDNC, and around $22\%$ over Loss-Unaware IDNC, which is significant. Furthermore, the average of constrained completion time over all frames is 2.5 packet transmissions in our simulation, while the average completion time in Loss-Aware IDNC and Loss-Unaware IDNC is 8.5 and 9.5 transmissions, respectively. \Ie Content-Aware IDNC improves by more than 70\% over IDNC in terms of delay.

Note that among the four methods of Content and Loss-Aware IDNC, Content-Aware Loss-Unaware IDNC, Loss-Aware IDNC, and Loss-Unaware IDNC, the performance of Content and Loss-Aware IDNC is the best and the performance of Loss-Unaware IDNC is the worst. The reason is that content-awareness and loss-awareness are the two components that we consider in our method, Content and Loss-Aware IDNC, to improve the performance of IDNC. Since Loss-Unaware IDNC does not consider either of these components, it has degraded performance. The two methods Content-Aware Loss-Unaware IDNC and Loss-Aware IDNC considers only one of these components; Content-Aware Loss-Unaware IDNC considers only content-awareness and Loss-Aware IDNC considers only loss-awareness. The wider the range of variation for probabilities of channel losses for D2D links in the local area, the more improvement is obtained by using loss-aware methods. On the other hand, the wider the range of variation for the importances of packets, the more improvement is obtained by using content-aware methods.

\begin{table}  % is used to refer this table in the text
\caption{Total Distortion Improvement of Content and Loss-Aware IDNC-P$_2$ over (i) Content-Aware Loss-Unaware IDNC-P$_2$, (ii) Loss-Aware IDNC, and (iii) Loss-Unaware IDNC} % title of Table
%\vspace{-10pt}
\centering % used for centering table
\begin{tabular}{c c c c} % centered columns (4 columns)
\hline
%\multicolumn{1}{c}{} & \multicolumn{3}{|c|}{Total Distortion}\\
%\hline\hline %inserts double horizontal lines
Video & (I) & (II) & (III)\\ [0.5ex]
\hline % inserts single horizontal line
Akiyo & $16\%$ & $9.3\%$ & $25.1\%$\\ % inserting body of the table
Grandma & $13\%$ & $5.4\%$ & $18.9\%$\\
%News & $19.5\%$ & $20.7\%$ & $28.8\%$ & $3.8\%$ & $4.6\%$ & $4.9\%$ % [1ex] adds vertical space
\end{tabular} \label{table:VT}
\vspace{-10pt}
\end{table}

{\bf Complexity:} We note that our optimization algorithms rely on finding cliques with maximum weights to determine the best network codes. This introduces complexity as clique finding problem is NP-complete. Yet, the complexity is not a bottleneck in our practical system setup as (i) we assume that the content is divided into blocks of packets, and we run our algorithms over these blocks, and (ii) we are interested in a micro-setup where a small number of devices cooperate to exchange packets.
%Even if the number of packets and devices increases, computationally efficient heuristics for maximum weighted clique selection problem could be used.

%We leave the development of computationally efficient algorithms as a future work.
%(14+13.5)/2=13.75~14
%(39+38)/2=38.5~39
%(26+27)/2=26.5~27
%(32+31)/2=31.5~32
%granma and akiyo had larger variance of priority values so better performance

%and also the minimum distortion among all users are 

\section{\label{sec:conc}Conclusion}
In this paper, we proposed a novel framework to improve the performance of instantly decodable network coding by exploring content-awareness. We considered a setup in which a group of mobile devices are interested in the same content, but each device has a partial content. Then, the devices cooperate by exploiting their D2D connections to receive the missing content. IDNC has been used to reduce the completion time when all devices receive the complete content. In practical applications, such as video streaming, not all packets have the same importance. In such applications, each device is interested in receiving a high quality content, instead of the complete content. We proposed Content and Loss-Aware IDNC that delivers a high quality content to each device by taking advantage of the contributions different parts have to the content. Simulation results showed significant improvement over baselines.

\end{document}